\documentclass[pdflatex,sn-nature]{sn-jnl}
\usepackage{caption}

\usepackage[utf8]{inputenc}
\usepackage[T1]{fontenc}
\usepackage{float}
\usepackage{siunitx}
\usepackage{relsize}
\usepackage{xspace}
\usepackage{algorithm}
\usepackage{algpseudocode}
\usepackage{graphicx}
\usepackage{amsmath}
\usepackage{bm}

\DeclareRobustCommand{\moldstruct}{%
  M{\protect\relsize{-2}OL}DS{\protect\relsize{-2}TRUCT}\xspace
}

\begin{document}

\title[Protein eXplosion Imaging]{Protein eXplosion Imaging (PXI): Protein Structures from Laser-Driven Explosions}

\author[1]{\fnm{Alfredo} \sur{Bellisario}}
\equalcont{These authors contributed equally to this work.}

\author[1]{\fnm{Tomas} \sur{Andr\'e}}
\equalcont{These authors contributed equally to this work.}

\author*[1,2]{\fnm{Carl} \sur{Caleman}}\email{carl.caleman@physics.uu.se}
 
\author*[1]{\fnm{Nicu\c{s}or} \sur{T\^\i mneanu}}\email{nicusor.timneanu@physics.uu.se}

\affil[1]{\orgdiv{Department of Physics and Astronomy}, \orgname{Uppsala University},
\orgaddress{\street{Box 524}, \postcode{SE-75120}, \city{Uppsala}, \country{Sweden}}}

\affil[2]{\orgname{Center for Free-Electron Laser Science, Deutsches Elektronen-Synchrotron},
\orgaddress{\street{Notkestra\ss e 85}, \postcode{DE-22607}, \city{Hamburg}, \country{Germany}}}

\abstract{%
We investigate the structural information retained in the distribution of explosion ion trajectories from proteins subjected to strong ionization. Using molecular-dynamics simulations and machine-learning analysis benchmarked against an analytical approach, we show that low-resolution structural information can be retrieved from the explosion distributions alone. 
An ensemble of convolutional neural networks trained on simulated spherical ion maps recovers the radius of gyration and the three semi-axes of an ellipsoidal molecular envelope with prediction errors of 1.2 {\AA} and 1.5 {\AA}, respectively, while an analytical ellipsoid charge model returns similar estimates and performs best for globular structures. We further investigate how higher-level structural information and symmetries can be extrapolated from ion measurements. This study supports the viability for structural determination of single proteins without large-scale X-ray facilities.
}

\raggedbottom
\maketitle

\section*{Introduction}

Laser-induced explosions occur when bound electrons are excited by the incident radiation, leading to the formation of positively charged ions; the resulting Coulombic repulsion between these ions breaks atomic bonds and generates a plasma explosion, a process known as Coulomb Explosion (CE). By collecting ions and recording their momenta, molecular structures can be determined, using a method called Coulomb Explosion Imaging (CEI)~\cite{kanter1979role, vager_coulomb_1989}. CEI has been utilized to investigate rare conformers and quantum fluctuations in small molecules~\cite{boll_x-ray_2022,Janke2025,boll2025,Wang2026}. However, for larger molecules, collecting momenta for multiple ions becomes increasingly challenging, and CEI is rarely applied to structures containing more than a dozen atoms. 

Theoretical studies suggest that the spatial positions of ions after explosion are reproducible for single atomic species in proteins \cite{ostlin2018} and that the root mean square error (RMSE) of the ion positions does not diverge dramatically even at different levels of laser fluence \cite{andre2026orientation}, directly linking the initial positions of the atoms to the final maps. For X-ray free-electron laser (XFEL) driven experiments, rapid inner-shell photoionization and subsequent Auger--Meitner cascades can create high charge states before the molecule undergoes extensive structural rearrangement, allowing the ensemble ion distribution to preserve coarse information about the initial geometry \cite{ostlin2018,dawod_moldstruct_2024,andre2026orientation}. 
\begin{figure}[H]
    \centering
    \includegraphics[width=1\linewidth]{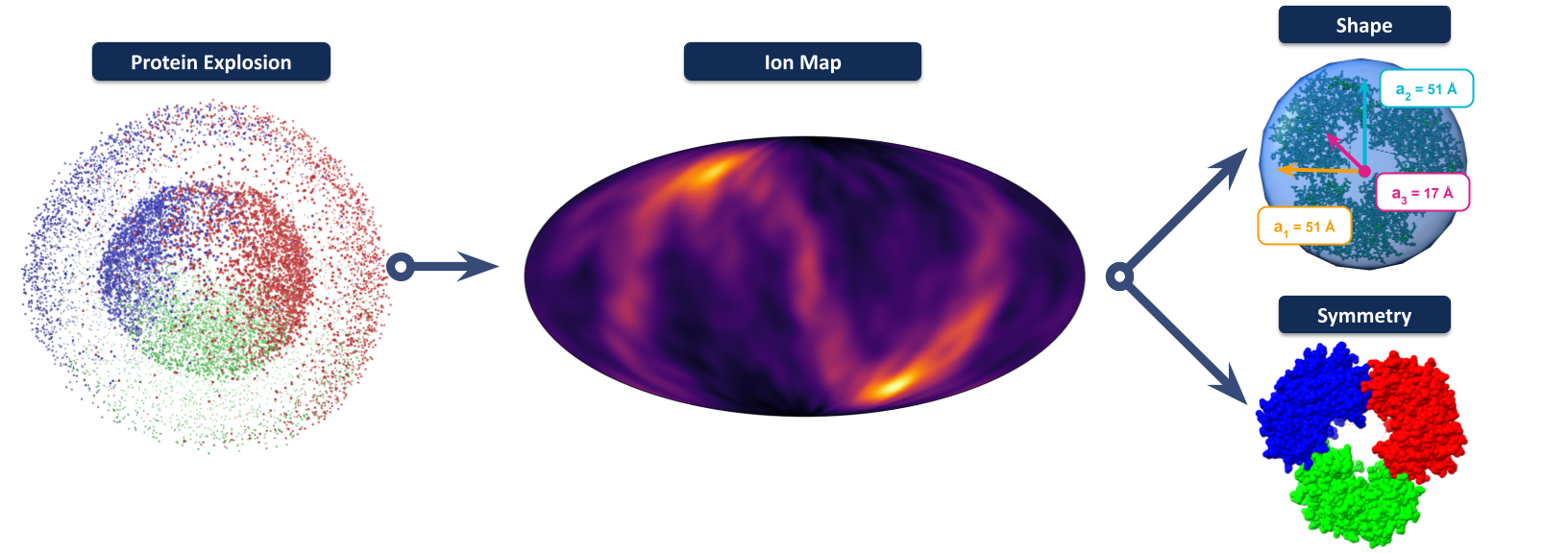}
    \caption{Schematic overview of the pipeline for inferring three-dimensional structural information from protein explosion maps, which we refer to as ion maps. A supervised model is trained to predict structural descriptors of proteins from a full \(4\pi\) explosion fingerprints, such as the radius of gyration \(R_g\) and the semi-axes \(a_1\), \(a_2\), \(a_3\). We benchmarked the predicted size parameters against analytical estimates. From these size parameters we can assemble an envelope of the original protein. We can also determine whether the ion distributions encode rotational symmetry, giving further insight into low-resolution structural features.}
    \label{fig:overview}
\end{figure}

When the ion signal is integrated over all atoms, the structural dependent ion trajectories enable the detection of very small conformational changes~\cite{andre2024}, such as those found in the $\beta$-hairpin loop of a viral capsid protein. Recent studies have shown that explosion footprints can be classified using unsupervised machine-learning algorithms, including dimensionality-reduction methods such as principal component analysis (PCA) and t-distributed stochastic neighbor embedding (t-SNE), as well as clustering~\cite{andre2024,AndrePartial2025}. Additionally, the relative orientation of a sample can be determined from ion measurements of an exploding protein structure~\cite{andre2026orientation}, which may significantly benefit Single-Particle Imaging (SPI) using coherent diffraction.

In this work, we introduce Protein eXplosion Imaging (PXI), a protein-scale analog of CEI that uses the \emph{spatial positions} of ions from ionized proteins to determine molecular size, shape, and symmetry. We refer to the time-integrated angular ion distribution as a PXI map or, equivalently, as an ion map. Rather than attempting an atomic-level reconstruction directly from the ion positions, here we show which structural information is retained and what can be learned a-priori from measurement of the emitted-ion distributions. For many structural biology applications, low-resolution descriptors, such as those from  Small Angle X-ray Scattering (SAXS) measurements, are already useful: radius of gyration, elongation, and folding can help distinguish conformational states, identify heterogeneous assemblies, and provide insights into the functional landscape accessible to proteins. Unlike a one-dimensional SAXS profile, a full angular ion map is a 2D observable that is strongly dependent on the original atom positions~\cite{andre2024}. 

We tested this idea using simulated Coulomb explosions from a protein dataset. A simulation framework (\moldstruct \cite{dawod_moldstruct_2024,kruger2026moldstruct}), using molecular dynamics simulations in \textsc{Gromacs}~\cite{Pronk2013}, was recently developed to investigate Coulomb explosion (CE) phenomena in proteins by modeling the complex photon-matter interactions that occur during XFEL exposure \cite{dawod_moldstruct_2024}. This framework enabled the generation of a dataset through molecular dynamics simulations that capture ionization dynamics and explosion signatures of various molecules under controlled conditions. As illustrated in FIG.~\ref{fig:overview}, the proposed workflow uses simulated ion maps and extracts complementary structural information, including molecular size, shape, and symmetry. {We will show that simulated ion maps retain measurable information about the molecular envelope, its size, and symmetries, which can be extracted with machine learning}. Further, we develop an analytical model, here called \emph{ellipsoid charged model}, based on first principles of electric fields in uniformly charged ellipsoids to directly calculate the same structural descriptors as the machine learning model. 

These results support the use of ion distributions as observables for low-resolution structural characterization of proteins subjected to intense ionizing pulses. In principle, Coulomb explosions need not necessarily be limited to XFEL facilities. We speculate that with strong-field optical tabletop lasers, structural retrieval is also possible when multiple ionization occurs within a sufficiently short pulse, although field-driven electron dynamics and geometric rearrangement during sequential ionization can introduce a stronger channel and intensity dependence \cite{ashrafi-belgabad_reconstructing_2024}. If sufficiently rapid ionization and ion detection can be achieved with optical lasers, PXI could potentially be implemented in a tabletop setup, making these measurements less costly and more widely accessible.

\section*{Results}

\begin{figure}[H]
    \centering
    \includegraphics[width=0.9\linewidth]{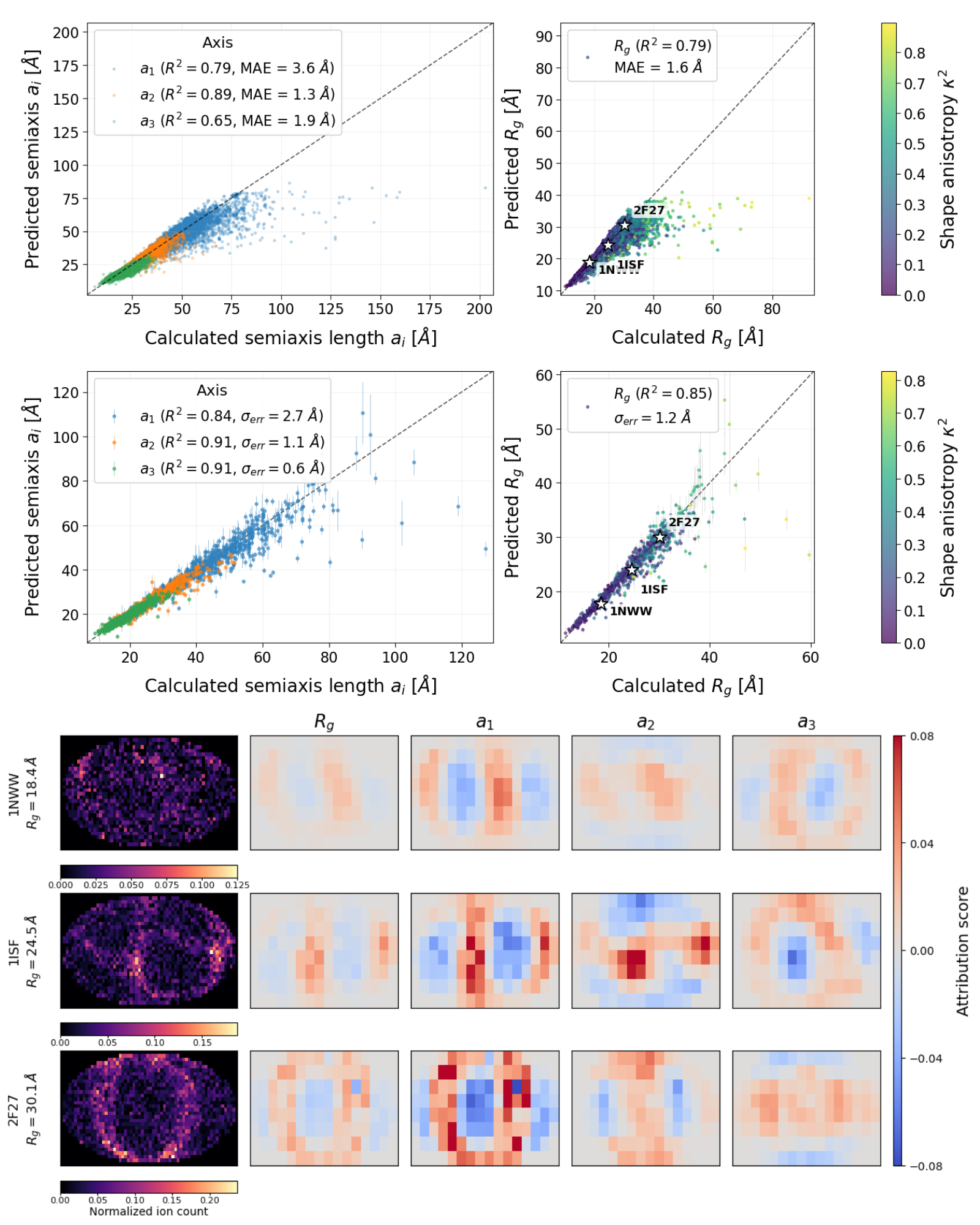}\\    
    \caption{Recovery of protein dimensions from ion maps. Top panel: analytical estimates of the three semi-axes, \(a_1,a_2,a_3\), and radius of gyration, \(R_g\), obtained directly from the explosion-map gyration tensor and compared with values calculated from the PDB protein coordinates over the entire dataset. Middle panel: corresponding machine-learning predictions over the test dataset. Dashed lines indicate perfect agreement, and points in the \(R_g\) plots are colored by the shape anisotropy \(\kappa^2\). Bottom panel: pixel-attribution analysis is shown for three representative test structures (PDB-IDs: 1NWW \cite{Arand2003_1NWW}, 1ISF \cite{YamamotoKatayama2002_1ISF}, 2F27 \cite{Chavas2006_2F27}). Attribution scores are computed as gradient \(\times\) input for each model and averaged across the model ensemble. Red regions contribute positively and blue regions negatively to each prediction.}
    \label{fig:predictions}
\end{figure}

\begin{figure}[H]
    \centering
    \includegraphics[width=\linewidth]{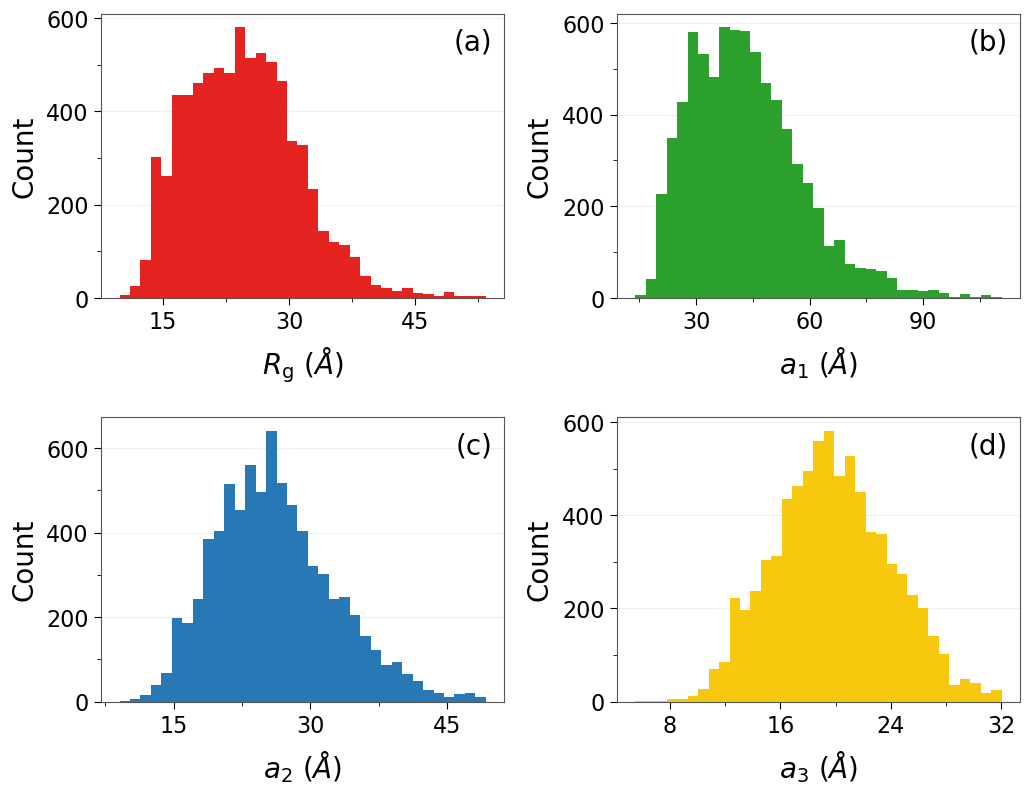}
    \caption{Distributions of protein size and shape descriptors across the dataset after outlier filtering: (a) radius of gyration $R_g$, and (b)--(d) the major, intermediate, and minor semi-axes $a_1$, $a_2$, and $a_3$, respectively. The histograms are shown up to the 99th percentile; the few remaining values above this range are not displayed.}
    \label{fig:distributions}
\end{figure}

The dataset initially comprised 7,644 distinct protein structures spanning a broad range of sizes and shapes, from compact, approximately globular proteins to elongated and multimeric assemblies. Of these, 85 PDB entries were excluded as described in the Methods, leaving 7,559 structures for the analysis reported here. Structural descriptors shown in figures were calculated using only $C_{\alpha}$ atoms from the backbone of the molecules. For each protein structure, we simulated one Coulomb-explosion trajectory and generated a corresponding base explosion map; random rotations of these maps were applied during model training at every epoch. We calculated structural descriptors directly from the atomic coordinates; the average radius of gyration $R_g$ in our dataset is \(24.5\pm6.7\)~\AA{}, while the distributions of the ellipsoidal semi-axes and  $R_g$ are shown in FIG.~\ref{fig:distributions}. The dataset spans a wide range of protein sizes, dimensions, and structural compositions. The structural diversity is summarized in Table~\ref{tab:shape_descriptors}, which reports the distributions of the size and shape descriptors. We set aside \(10\%\) of the structures as an independent test set, while the rest were used for repeated cross-validation and ensemble training.

Because Coulomb-driven redistribution is anisotropic, the relative dimensions of the protein remain reflected in the eigenvalues of the ion-distribution gyration tensor. We first tested whether global molecular dimensions could be inferred from the explosion maps without using machine learning. Experimentally, ions emitted from the interaction region can be recorded on a spatially resolved multi-channel plate (MCP) detector. The recorded ion-hit positions form a planar two-dimensional explosion footprint, which can be converted into emission directions using the known detector geometry and projected onto the unit sphere. Although a finite detector captures only part of the angular distribution, the complete $4\pi$ distribution can be reconstructed from partial measurements~\cite{andre2026orientation}; here, we therefore consider complete $4\pi$ angular coverage. The eigenvalues of the gyration tensor of this angular distribution can be related to those of the original molecular structure. Using the eigenvalues of the unweighted gyration tensor of each ion map, we estimated the semi-axes and $R_g$ with the ellipsoid charged model described in Methods on the whole dataset and compared these estimates with values calculated directly from the PDB coordinates (FIG.~\ref{fig:predictions}). 

The analytical estimates are correlated with the true dimensions of the proteins, with $R^2=0.79$, $0.89$, and $0.65$ for $a_1$, $a_2$, and $a_3$, respectively, and with corresponding mean absolute errors (MAEs) of $3.6$, $1.3$, and $1.9$~\AA{}. For $R_g$, these estimates give $R^2=0.79$ and an MAE of $1.6$~\AA{}. The largest deviations occur for elongated and strongly anisotropic structures, particularly along the longest semi-axis. This limitation can be anticipated directly from the analytical estimates. Since the estimated semi-axes can also be used to calculate the relative shape anisotropy \(\kappa^2\), the ellipsoid charged model provides not only an estimate of molecular size and shape, but also an indication of when that estimate is likely to be unreliable. Structures with low \(\kappa^2\) are generally described well by the analytical approximation, whereas the agreement deteriorates systematically toward high \(\kappa^2\). Thus, \(\kappa^2\) provides a useful internal indicator of the expected quality of our estimates.

We then trained a ensemble of convolutional neural networks directly on the spherical ion-density maps. All structures were simulated under identical XFEL conditions and random rotations were applied during training after each epoch to prevent the network from associating structural descriptors with a fixed orientation. Prediction errors are reported on the test dataset as the residual standard deviation $\sigma_{\mathrm{err}}$, in \AA{} for $R_g$ and the axes, together with the dimensionless coefficient of determination $R^2$. For the independent test set, the network predicted $R_g$ with $\sigma_{\mathrm{err}}=1.2$~\AA{} and $R^2=0.85$, which means that approximately $85\%$ of the variance in the true values of $R_g$ is explained by the model. For the semi-axes, the network achieved $R^2=0.84$, $0.91$, and $0.91$ for $a_1$, $a_2$, and $a_3$, with corresponding $\sigma_{\mathrm{err}}$ values of $2.7$, $1.1$, and $0.6$~\AA{}. The average $\sigma_{\mathrm{err}}$ across the three axes is therefore $1.5$~\AA{}. The predictions closely follow the identity line across most of the size range and substantially reduce the shape-dependent bias observed for the analytical estimates. The largest remaining errors occur for the longest semi-axis and for the most extended proteins, which represent only a small fraction of the dataset.

To identify which regions of an explosion map contribute to each prediction, we performed a pixel-attribution analysis on downsampled network inputs, with representative examples shown in FIG.~\ref{fig:predictions}. The attribution patterns for \(R_g\) and \(a_1\) show some common spatial features, consistent with both quantities being related to the spatial extent of the ion distribution. Since the principal directions associated with \(a_1\), \(a_2\), and \(a_3\) are mutually orthogonal, we asked whether orthogonality is also reflected in the information extracted by the network. To quantify this, for each protein in the test set we flattened the attribution maps into vectors and calculated the pairwise cosine similarity between the three semiaxis attributions, where values close to zero indicate roughly orthogonal attribution patterns and positive or negative values indicate alignment or anti-alignment, respectively. The cosine similarities are \(0.134 \pm 0.288\) (median \(0.127\)) for \(a_1\)--\(a_2\), \(-0.375 \pm 0.198\) (median \(-0.397\)) for \(a_1\)--\(a_3\), and \(-0.248 \pm 0.365\) (median \(-0.328\)) for \(a_2\)--\(a_3\). This weak alignment is consistent with the expectation of a rough orthogonality between the information used for the three principal dimensions. Much of the map can contribute to all three predictions, with differences in sign and relative importance. The observed cosine similarities therefore suggest that the network extracts overlapping spatial information while organizing and weighting it differently for each principal dimension. This behavior is also consistent with the expected effect of rotational augmentation during training, which encourages the model to treat each principal direction independently of a fixed image orientation and instead learn orientation-robust structural features.

\begin{figure}[H]
    \centering
    \includegraphics[width=\textwidth]{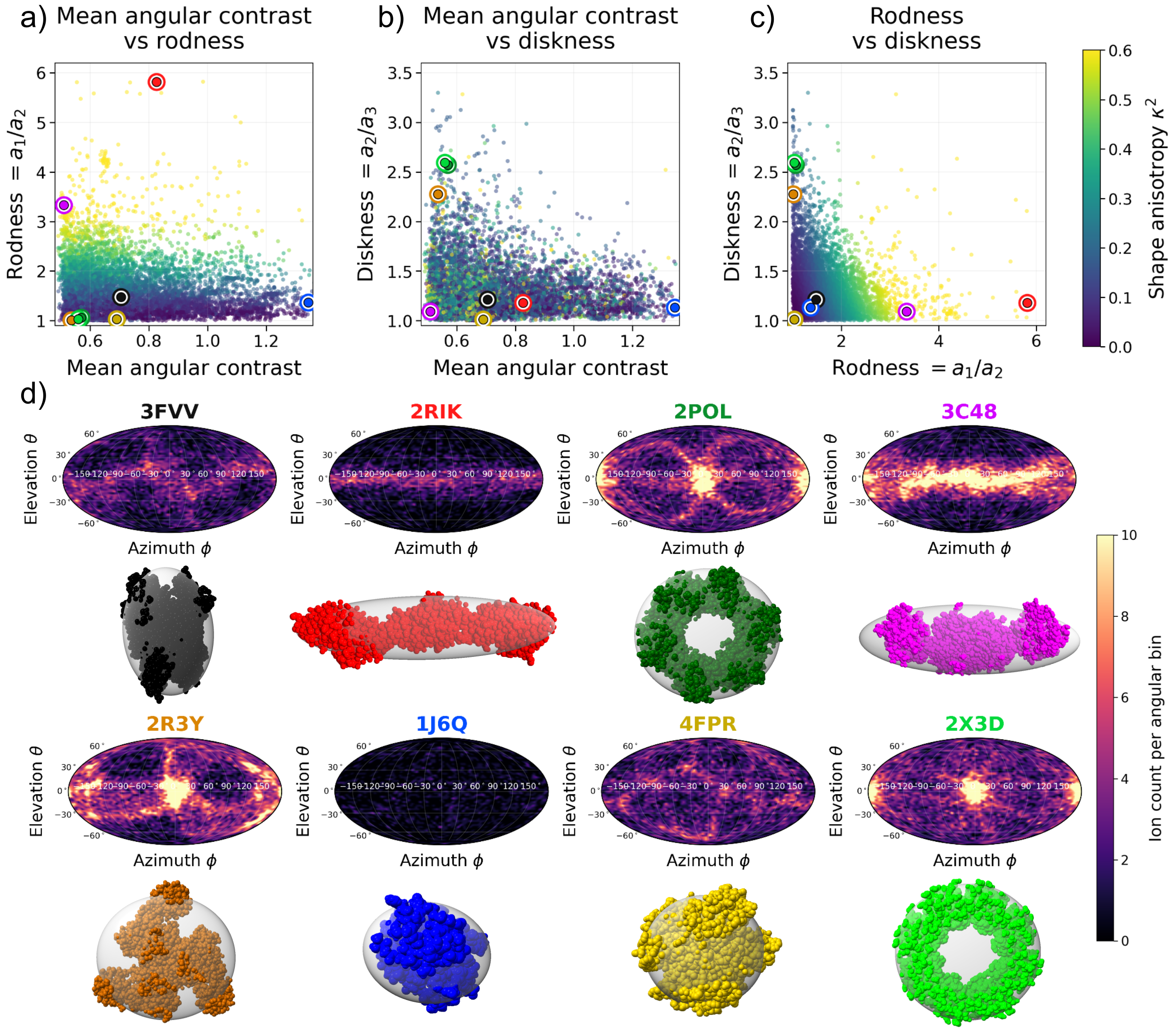}
    \caption{Top panels: protein-shape space derived from metrics computable directly from ion maps. Each point represents a single protein, positioned according to three shape descriptors: the mean angular contrast of its corresponding projected ion map and two semi-axis ratios derived from the ellipsoid axes. We define the ratio \(a_1/a_2\) as rodness, which quantifies elongation, and the ratio \(a_2/a_3\) as diskness, which quantifies flattening along the shortest axis. Points are colored by the relative shape anisotropy \(\kappa^2\). A small number of proteins with extreme values outside the displayed shape-space region were excluded from the visualization. Bottom panel: eight selected examples are highlighted across the shape-space plots; the lower panels show their ion maps and native structures. Structures shown have PDB-ID: 3FVV~\cite{Zhang2009_3FVV}, 2RIK~\cite{doi:10.1073/pnas.0707163105}, 2POL~\cite{KONG1992425}, 3C48~\cite{VETTING200815834}, 2R3Y~\cite{hasselblatt2007regulation}, 1J6Q~\cite{10.1021/bi026362w}, 4FPR~\cite{https://doi.org/10.1111/tpj.12913}, and 2X3D~\cite{Oke2010}. For the selected examples, fitted ellipsoids derived from the ensemble-averaged ML predictions are overlaid on the PDB structures to illustrate how the predicted semi-axis dimensions capture the overall protein shape. The protein densities are not to scale and are not in the same frame as the ion maps.}
    \label{fig:shape-space}
\end{figure}

The directional organization observed in the attribution maps suggests that the network predictions arise from spatial variations in the ion map, as seen in FIG.~\ref{fig:predictions}. To quantify spatial variations directly on the sphere, we calculated the mean angular contrast of each ion-density map from the absolute differences in normalized ion density between neighboring HEALPix pixels. After normalizing each ion-density map to unit total intensity, these differences were averaged over all unique neighboring pixel pairs and normalized by the mean pixel density. This provides a dimensionless measure of how strongly the ion density varies between adjacent emission directions on the sphere. We then constructed a low-dimensional shape space using the mean angular contrast of the ion distribution together with the two semi-axis ratios and the shape anisotropy. The resulting distribution is shown in FIG.~\ref{fig:shape-space}. The relative shape anisotropy, $\kappa^2$, indicated by the color scale, quantifies the deviation of a structure from a sphere and is calculated from the eigenvalues of the gyration tensor. It therefore complements the semi-axis ratios by providing a single dimensionless measure of the overall anisotropy. Here, rodness is defined by the ratio $a_1/a_2$ and describes the degree to which the longest semi-axis exceeds the intermediate one, whereas diskness is defined by $a_2/a_3$ and describes the degree to which the intermediate semi-axis exceeds the shortest one. Proteins with nearly equal semi-axes cluster near low rodness and diskness values and have low $\kappa^2$, whereas increasingly anisotropic structures extend toward larger axis ratios and higher $\kappa^2$. Rod-like structures extend toward high $a_1/a_2$, corresponding to increasing elongation along one axis, while disk- or slab-like structures are characterized by a larger $a_2/a_3$ ratio, corresponding to increasing flattening along the shortest axis.

The selected structures in FIG.~\ref{fig:shape-space} show that their positions in this space correspond to differences in global shape. The axis ratios account for the dominant organization of this space, while the mean angular contrast captures additional variation among structures with similar principal-axis ratios. For example, 3FVV and 1J6Q occupy relatively similar positions in the rodness--diskness plane but differ substantially in their mean angular contrast. 
More generally, the mean angular contrast exhibits a relationship with the range of accessible global protein shapes. Strongly rod- or disk-like structures occur predominantly at low angular contrast, whereas high angular contrast is largely confined to proteins with rodness and diskness values close to unity. The converse is not true: low angular contrast spans a broad range of rodness and diskness, indicating that the scalar angular contrast constrains, but does not uniquely determine, the global protein geometry. Compact, elongated, toroidal, and flattened representatives therefore sample different regions of this shape distribution. The relationship between the angular variation of the explosion map and independently calculated global shape descriptors indicates that the ion map retains information about molecular geometry. 

\begin{figure}[H]
    \centering
    \includegraphics[width=\linewidth]{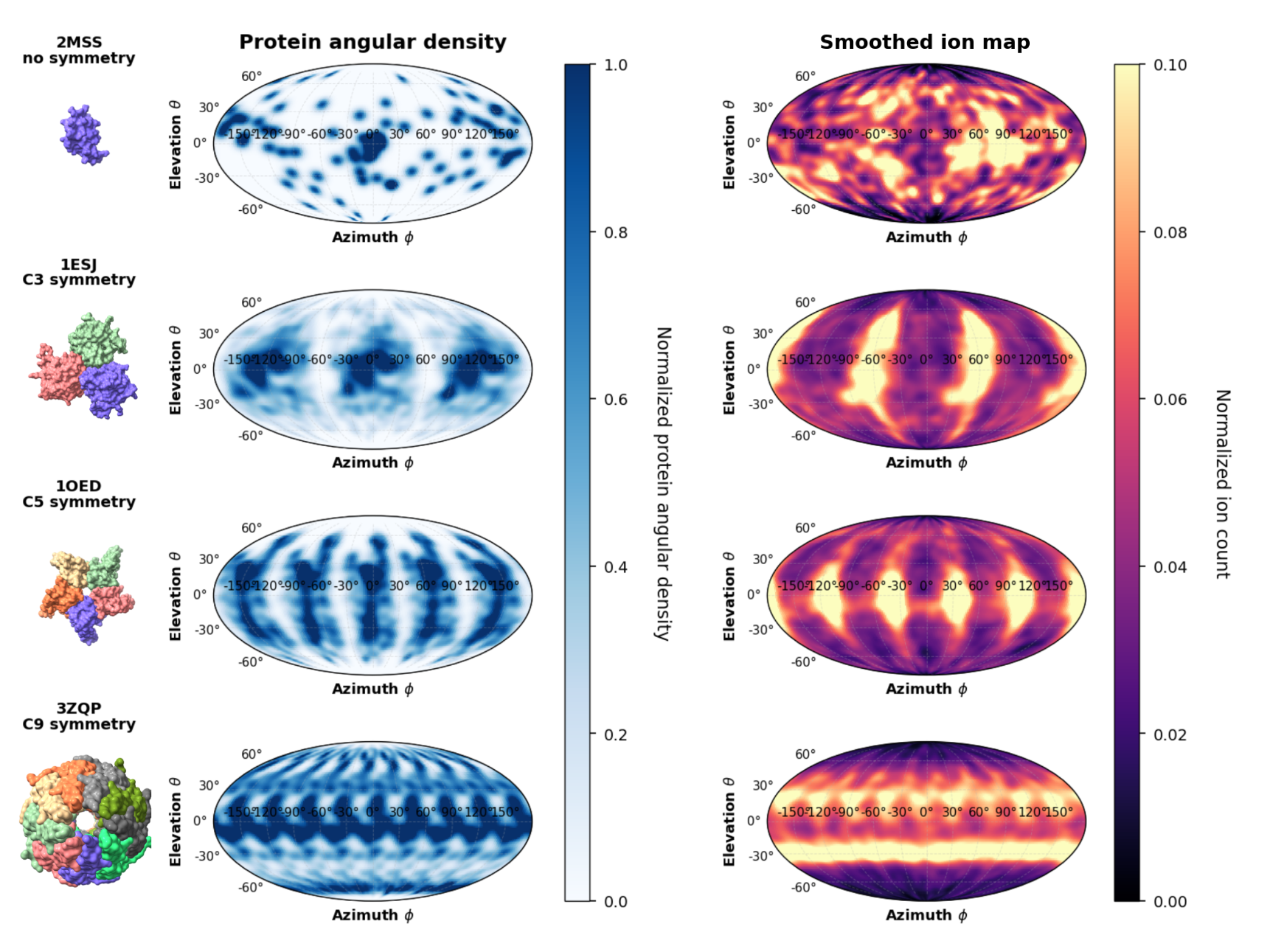}
    \caption{%
    Angular protein density and ion maps for four chosen protein assemblies. Left: Protein structures with different rotational symmetries - 2MSS (no rotational symmetry)~\cite{Nagata1999Musashi}, 1ESJ ($C_3$)~\cite{Campobasso2000ThiK}, 1OED ($C_5$)~\cite{Miyazawa2003AChR}, and 3ZQP ($C_9$)~\cite{Buttner2012Terminase}. Center: Mollweide projections of the normalized angular distributions of the protein C$_{\alpha}$ coordinates. Right: Mollweide projections of the corresponding ion map after the Coulomb explosion, using exactly the same symmetry-alignment rotation as the protein.
    }
    \label{fig:symmetries}
\end{figure}

The separation of proteins with similar ellipsoidal dimensions suggests that the ion maps also encode higher-order angular organization beyond global size and axis ratios. To probe this information using a well-defined structural feature, we show the preservation of rotational symmetry in the ion maps and compare it to the angular distribution of the electron density. FIG.~\ref{fig:symmetries} compares the electron density and the ion maps for four proteins spanning different rotational symmetries, from 2MSS with no rotational symmetry to 1ESJ, 1OED, and 3ZQP with $C_3$, $C_5$, and $C_9$ symmetry, respectively. For each $C_n$ structure, the rotational axis that maximizes the corresponding $C_n$ symmetry of the protein angular density is aligned with the $+z$ direction before the angular distribution is calculated. The protein angular distributions and the ion maps are Gaussian-smoothed with widths $\sigma_{\theta}=4^\circ$ and $\sigma_{\phi}=6^\circ$ in the polar and azimuthal directions, respectively. The angular density distribution of the protein is calculated from the rotated coordinates relative to the geometric centroid of the $C_{\alpha}$ atoms and represented by a Mollweide projection. During the explosion process, each ion is accelerated by the local electric field generated by the surrounding charges. In the examples shown here, this redistribution produces an approximate \emph{anticorrelation} between the protein angular distribution and the ion emission pattern, such that directions with a higher densities often correspond to directions with less ions, and vice versa, as seen in FIG.~\ref{fig:symmetries}. Furthermore, the characteristic rotational organization remains visible in the ion maps: the number and angular spacing of repeated features follows the underlying $C_n$ symmetry, while the 2MSS structure shows the corresponding nonperiodic distribution. This shows that the emitted-ion distributions retain coarse-grained angular information that distinguishes rotational $C_n$ symmetries across structurally distinct proteins.

\section*{Discussion and Conclusions}

Our results show that PXI observables can provide a complementary route to conventional imaging methods. Previous work demonstrated that PXI maps can classify closely related protein structures, even when they differ only in the conformations of a small number of amino acids \cite{andre2024}. Our findings indicate that the overall dimensions, anisotropy, and assembly symmetry of the initial protein are preserved on the explosion map. These quantities can be estimated by either direct analysis of the ion distribution or machine-learning regression. PXI thus can be used not only for structure classification but also for quantitative predictions. However, the attainable experimental resolution and the range of structures for which these relationships remain valid have yet to be determined. Figure~\ref{fig:shape-space} further shows that proteins with nearly identical semi-axis ratios can produce ion maps with markedly different angular contrast.

For future work, machine learning will play an important role in addressing the inverse problem associated with PXI. The ensemble training approach improved robustness to variation between structures and provided a direct estimate of predictive dispersion. We can imagine that these results can further be improved by fitting bead models into ion measurements, as already done for SAXS data, or by taking advantage of the correlation between ion maps and angular density. 

Integrating ion data collection in SPI measurements could also be an interesting approach in helping to develop gas-phase experiments at XFELs and improve the robustness of 3D reconstructions, especially when combined with traditional diffraction data. In practice, these ion-derived observables could be used to pre-classify particles into broad conformational groups before more computationally intensive reconstruction with SPI, or to provide physical constraints during orientation recovery in phasing of diffraction data. For example, for small proteins where the hit-ratio is very low and for out-of-focus measurements with very low signal-to-noise ratios, ion data collection can provide a rapid screening to identify real hits as already shown in literature~\cite{pietrini2018statistical}. With an experimental setup similar to the one described here and elsewhere~\cite{andre2024,AndrePartial2025,andre2026orientation}, we imagine that a rapid screening using axis ratios or estimated sizes could flag outlier states or identify mixtures of folded and unfolded conformations in heterogeneous samples. 

Several challenges remain in the present study that provide possible directions for future work. First, our framework relies on simulated explosion trajectories under idealized detector conditions. Real experimental ion maps will include additional sources of uncertainty, such as detector inefficiencies, incomplete angular coverage, ion losses, timing jitter, and electronic and background noise. The robustness of PXI to several of these effects has already been investigated in the literature \cite{andre2026orientation} (in Supplementary Information), providing a basis for extending the present framework to experimental data. Second, this analysis focused only on descriptors that capture molecular size and symmetry. Although these descriptors define the boundaries of the molecular envelope, they do not provide a unique atomic structure. Electron density features and specific secondary-structure motifs were not addressed in this study. Third, all models were developed using a relatively small set of protein data, each simulated under the same explosion conditions. It remains to be established how well the learned structure--ion relationships generalize across different ionization and experimental conditions.

A broader implication of this and previous work \cite{andre2024} is that Coulomb-driven structural imaging does not necessarily need to remain tied to large-scale XFEL facilities. In principle, if sufficiently controlled ionization and ion-detection schemes can be implemented using optical or strong-field laser systems, similar structural observables may become experimentally accessible in much smaller and more widely available setups - such as tabletop experiments. Experimentally, the main challenge is not only generating the explosion, but also accurately detecting the resulting ions with enough angular coverage and single-shot resolution without detector saturation. 

\section*{Methods}

\subsection*{A. Coulomb Explosion Simulations and Ion Maps}

\subsubsection*{Coulomb Explosion Simulations}

All ionization and Coulomb-explosion simulations were performed using \moldstruct~\cite{dawod_moldstruct_2024,kruger2026moldstruct}, a Monte Carlo/Molecular Dynamics framework implemented in \textsc{Gromacs}~4.5.6 with the CHARMM36 force field. Protein structures were first energy minimized using steepest descent and equilibrated in vacuum at 300~K using a timestep of 1~fs while maintaining fixed molecular orientations. For each protein, equilibrated conformations were extracted and used as starting points for independent ionization simulations. A single explosion simulation was generated per structure. The ionization model includes photo-ionization, fluorescence emission, and Auger--Meitner decay processes. Consistent with the system sizes considered here, all liberated electrons were assumed to escape the system. The initial center-of-mass translational velocity associated with particle injection into the interaction region ($10^1$--$10^2$~m/s) was neglected and set to zero. This approximation is valid because the injection velocity is several orders of magnitude smaller than the characteristic Coulomb-explosion velocities ($10^4$--$10^5$~m/s), and therefore contributes only a negligible center-of-mass drift. The explosion dynamics was propagated with a time step of 1~as. The XFEL pulse was modeled as a Gaussian pulse centered at $t=20$~fs with a full width at half maximum (FWHM) of 10~fs, photon energy of 2~keV, and fluence of $5\times10^{6}$ photons/nm$^2$, corresponding to experimental conditions at the SQS instrument of the European XFEL~\cite{meyer2022sqs}. Each simulation was propagated until the total energy of the system was 99\% or more kinetic energy.

\vspace{0.5em}

\subsubsection*{Spherical Projections}

Coulomb-explosion simulations were generated for a dataset comprising 7644 protein structures obtained from the Protein Data Bank (PDB) \cite{Berman2000PDB}. One simulation was performed per PDB-ID and each explosion map was generated once. Detector hit coordinates were mapped onto the unit sphere and converted into spherical ion maps using the \textit{Hierarchical Equal Area isoLatitude Pixelisation} (HEALPix) framework~\cite{Gorski_2005,Zonca2019}, with $\mathrm{NSIDE}=64$, corresponding to 49,152 equal-area pixels and a characteristic angular scale of approximately $0.9^\circ$.

\vspace{0.5em}

\subsection*{B. Machine Learning}

\subsubsection*{Dataset}

\begin{table*}[t]
\caption{Shape descriptors computed from the gyration-tensor eigenvalues
for the processed dataset of Protein Data Bank (PDB) structures.
The radius of gyration $R_g$ measures overall compactness, $b$ the
deviation from spherical shape, $c$ the deviation from cylindrical
symmetry, and $\kappa^2$ the relative shape anisotropy. The axes
$a_1$, $a_2$, and $a_3$ describe the major, intermediate, and minor semi-axes of the
molecular envelope.}
\label{tab:shape_descriptors}
\centering
\renewcommand{\arraystretch}{1.12}
\begin{tabular*}{\textwidth}{@{\extracolsep{\fill}}lcc@{}}
\hline
Metric & Min & Max \\
\hline
Radius of gyration ($R_g$, \AA)
    & 9.9 & 91.9 \\
Asphericity ($b$, \AA$^2$)
    & 3 & 8018 \\
Acylindricity ($c$, \AA$^2$)
    & 0 & 745 \\
Shape anisotropy ($\kappa^2$)
    & 0.0 & 0.9 \\
Major semi-axis ($a_1$, \AA)
    & 14.0 & 202.0 \\
Intermediate semi-axis ($a_2$, \AA)
    & 9.1 & 67.0 \\
Minor semi-axis ($a_3$, \AA)
    & 5.5 & 38.3 \\
\hline
\end{tabular*}
\end{table*}

In the present work, the machine-learning models were trained directly on 2D projections of the $4\pi$ steradian spherical ion maps that represent the complete angular emission pattern of the explosion. During training, random orientations for each protein were generated at each epoch to account for random alignment in the laboratory frame. Each spherical ion map was paired with a corresponding vector of geometric descriptors derived from the underlying protein structure. The regression targets considered here were the radius of gyration and the semiaxes of the ellipsoidal envelope of each structure. The complete dataset was randomly shuffled using a fixed seed for reproducibility. A 10\% subset of structures was kept aside as an independent test set, while the remaining 90\% were used for repeated cross-validation training.

For each protein structure, shape descriptors were computed directly from the atomic coordinates using the gyration tensor formalism. The gyration tensor eigenvalues $(\lambda_1,\lambda_2,\lambda_3)$ were used to derive global shape descriptors that include the radius of gyration ($R_g$), asphericity, acylindricity, relative shape anisotropy,and semiaxes ratios. Structures with subunits separated by 2~nm or more, structures for which the Coulomb-explosion simulation failed, and structures whose elongation exceeded the range considered in this work were excluded from the machine-learning predictions and analytical estimates.

\vspace{0.5em}

\subsubsection*{Neural Network model,  training, \& interpretability}

We used a residual convolutional neural network for regression on spherical ion maps. Input maps consisted of single-channel HEALPix projections represented as two-dimensional tensors. The network architecture consisted of an initial convolutional stem followed by a hierarchy of residual convolutional blocks with progressively increasing channel depth (32, 64, 128, and 256 filters). Each residual block contained two convolutional layers with batch normalization and Swish activations~\cite{ramachandran2017searching}, providing a smooth nonlinear activation throughout the network. Spatial downsampling was performed using stride-2 convolutions. Global average pooling and global max pooling were applied to the final feature maps and concatenated into a shared latent representation. Afterward, fully connected layers of sizes 256 and 128 were used prior to the final regression output layer. The network was optimized using Adam optimizer~\cite{kingma2015adam} with a learning rate of $3 \times 10^{-4}$ and Huber loss function~\cite{huber1964robust}. To improve robustness and enable uncertainty estimation, models were trained as an ensemble using repeated five-fold cross-validation~\cite{dietterich1998approximate}. Specifically, four independent repeats of five-fold cross-validation were performed using different random seeds, resulting in a total of 20 independently trained models. For each fold, the training data was augmented by randomly sampling the orientation of each protein, whereas the validation and test datasets were kept fixed and were not augmented. Model training used early stopping based on validation loss to reduce overfitting~\cite{prechelt}. Learning rates were adaptively reduced using a plateau scheduler. For every member of the ensemble, both the final trained model and the best validation-loss checkpoint were stored independently. The final performance of the model was evaluated on the test set. 

For model interpretability, pixel-wise attribution maps were calculated using the gradient \(\times\) input method~\cite{ancona2018towards}. For each predicted structural descriptor, the gradient of the network output with respect to each input pixel was evaluated and multiplied by the corresponding normalized pixel intensity, \(A_i = x_i\,\partial y/\partial x_i\). Gradients were evaluated after rescaling the network output to the physical target scale, and the attribution maps were averaged across the independently trained models in the ensemble. The resulting signed attribution values quantify the local sensitivity of a prediction to the observed signal at each pixel, with positive and negative values corresponding to contributions associated with increasing and decreasing the predicted quantity, respectively. No normalization was applied to the attribution scores, and the maps were spatially coarse-grained by block averaging.

\section*{C. Ellipsoid charge model}
\subsection*{C1. Forward propagation --  from protein to explosion map}
Understanding the dynamics of the induced explosion gives insight into how to approach the inverse problem. A somewhat counterintuitive feature is that the angular distribution of the atoms is inverted between the pre- and post-explosion states: high-density regions in the angular density map onto low-density regions in the ion map, and vice versa (FIG. \ref{fig:symmetries}). Consider a rod-shaped protein with most of its atoms along the long axis. This is precisely the axis along which the fewest ions are ejected, since they instead leave perpendicular to it. The intuition is that ions escape the most highly charged regions as fast as they can: once the protein is ionized, each ion follows the electric field set up by the closest surrounding charges, which points away from the dense, strongly charged interior. We model this process by treating our charged protein as a uniformly charged ellipsoid, in which the electric field components \(E_i\) can be described with
\begin{equation}
    E_i \propto n_i x_i,
\end{equation}
where \(x\) is the position relative to the ellipsoid principal frame and \(n_i\) are the depolarization factors defined as
\begin{equation}
    n_i = \frac{a_1a_2a_3}{2}\int_0^{\infty} \frac{ds}{(s+a_i^2)\sqrt{(s+a_1^2)(s+a_2^2)(s+a_3^2)}},
\end{equation}
where \(a_i\) are the semi-axes of the ellipsoid such that \(a_1 \geq a_2 \geq a_3 \)~\cite{Bohren1998}. The factors \(n_i\) are scale independent and \(\sum n_i = 1\). This integral has no general closed form solution but can be evaluated numerically, and can be restated in terms of the Carlson symmetric form of elliptic integrals $R_D$ as
\begin{equation}
    n_i = \frac{a_1a_2a_3}{3}R_D(a_j^2,a_k^2,a_i^2),
    \label{eq:dep}
\end{equation}
which has numerically stable and accurate solvers. 

With the depolarization factors we can estimate the trajectory of each ion in the sudden approximation, where the ion acquires its momentum before the charge distribution
has appreciably changed. The ions direction is then set by the field at the
ion's initial position, proportional to\( (n_1x_1,n_2x_2,n_3x_3)\), so the map from initial
position to final direction is a linear stretch along the principal axes. We write
this stretch as
\begin{equation}
    A(\gamma) = \begin{pmatrix}
    n_1^\gamma & 0 & 0 \\
    0 & n_2^\gamma & 0 \\
    0 & 0 & n_3^\gamma
    \end{pmatrix},
    \label{eq:proj}
\end{equation}
where the exponent \(\gamma\) is set to 1 for now and we return to it below. Since
\(n_i\) depends only on the axes ratios, the map is invariant to the overall size of
the protein.

The final ion direction is then given by 
\begin{equation}
    v_{i,\mathrm{rec}} = \frac{Av_{i,0}}{|Av_{i,0}|},
    \label{eq:vrec}
\end{equation}
where \(v_{i,0}\) is the initial direction of the ion with regards to the ellipsoid center. By this method we can essentially reconstruct the ion map using \(v_{i,0}\) approximately by sampling \(v_{i,0}\) uniformly on the ellipsoid surface.

This allows us to reconstruct the gyration tensor as 
\begin{equation}
    M_{\mathrm{rec}} = \frac{1}{N}\sum_{i=1}^N v_{i,\mathrm{rec}}\cdot v_{i,\mathrm{rec}}^T.
    \label{eq:Mrec}
\end{equation}
This however assumes that the explosion of the protein is instantaneous, whereas in reality it is inherently a dynamical process, making the depolarization factors time-dependent due to the uneven expansion of the ellipsoid. This process adds additional anisotropy that is not accounted for by this forward model, and here \(\gamma\) will act as an anisotropy parameter, with \(\gamma < 1\) reducing anisotropy, and \(\gamma > 1\) increasing anisotropy. By calibrating on our dataset, we found that $\gamma = 1.25$ provides the optimal anisotropy correction. Across dataset splits and cross-validation, the optimal values of $\gamma$ lie approximately in the interval $[1.2, 1.3]$. Lower values describe small or elongated proteins, whereas higher values describe large or globular proteins. While randomized splits of the dataset yield essentially the same value of $\gamma$. Thus, although $\gamma$ does not appear to be strictly universal, it remains within a relatively narrow interval, and $\gamma = 1.25$ works well for the proteins in our dataset.

\subsection*{C2. Inverse problem -- how we can extract shape}
Using the tools introduced above we present a scheme that can extract the semi-axes and radius of gyration of the protein based only on the ion maps. We do this in seven steps:
\begin{enumerate}
    \item We build the gyration tensor, \(M_\mathrm{obs}\), from the observed ion vectors. Its eigenvectors give the orientation, and diagonalizing it transforms the tensor into the principal frame.

    \item Fix \(r_1=1\) and guess the ratios \(r_2 = a_2/a_1\),\(r_3 = a_3/a_1\).

    \item Run the forward step starting from a set of evenly sampled points on  the ellipsoid defined by \(a_1,a_2,a_3\) as described by Eq. \eqref{eq:dep}-\eqref{eq:Mrec}.

    \item Compute a cost function as \begin{equation}
        \mathrm{cost}(r_2,r_3) = \sum_{i=1}^3 (\mu_i^\mathrm{rec} - \mu_i^\mathrm{obs})^2,
        \label{eq:cost}
    \end{equation}
    where \(\mu^\mathrm{rec}\),\(\mu^\mathrm{obs}\) are the eigenvalues of the reconstructed and observed gyration tensor respectively. 

    \item Minimize eq. \eqref{eq:cost} using the Nelder-Mead optimizer~\cite{Nelder-Mean} to find optimal \(r_2,r_3\).

    \item Repeat steps 3–5 until the tolerance is reached (\(10^{-5}\) in the cost function); the median number of evaluations across the dataset is 60 and tolerance is reached for all proteins in less than 200 iterations. 
    \item Calculate the final eigenvalues, \begin{equation}
    \mu_i = \frac{r_i^2}{r_1^2 + r_2^2 + r_3^2},
    \label{eq:mu_from_r}
\end{equation}
from the two optimized ratios.
\end{enumerate}

This procedure gives the shape of the fitted ellipsoid, but no scale as \(n_i\) is scale invariant. To scale to physical units we need an estimate of how many atoms are in the protein, which we assume to be known in the experiment. We use the fact that the radius of gyration \(R_g\), is related to the scaled eigenvalues as \(R_g^2 = \sum \lambda_i\), where \(\lambda_i\) are the eigenvalues in the physical scale. The \(\lambda_i\) are related to the semi-axes \(a_i\) as 
\begin{equation}
    a_i = \sqrt{5\lambda_i} = R_g\sqrt{5\mu_i}.
    \label{eq:semiax}
\end{equation}

The volume of a protein scales with the number of atoms 
\begin{equation}
    V_\mathrm{atoms} = v_0N,
    \label{eq:vol_atoms}
\end{equation}
where \(v_0\) is the average volume per atom, based on the average protein density (including hydrogen) \(\rho = 1.37\)~g/cm$^3$~\cite{erickson2009size}. 
We can also calculate the volume of an ellipsoid based on the axes, related to our estimate eigenvalues \(\mu_i\) by Eq. \ref{eq:semiax} as
\begin{equation}
    V_\mathrm{ellipsoid} = \frac{4\pi}{3}a_1a_2a_3 = \frac{4\pi}{3}5^{3/2}R_g^3\sqrt{\mu_1\mu_2\mu_3}.
    \label{eq:vol_ell}
\end{equation}
Combining Eq. \eqref{eq:vol_atoms} and \eqref{eq:vol_ell} we can solve for the radius of gyration
\begin{equation}
    R_g = \Bigl(\tfrac{3v_0}{4\pi\,5^{3/2}}\Bigr)^{1/3} N^{1/3}
        \bigl(\mu_1\mu_2\mu_3\bigr)^{-1/6}.
        \label{eq:rg1}
\end{equation}
However, since this slightly underestimates the actual \(R_g\), arising from proteins fundamentally not being uniform ellipsoids, we can estimate how large this discrepancy is by directly calculating the volume of a protein

\begin{equation}
    f_V = \frac{V_\mathrm{ellipsoid}}{V_\mathrm{atoms}} = \frac{(4\pi/3)5^{3/2}\bar{R}_g^3\sqrt{\bar{\mu}_1\bar{\mu}_2\bar{\mu}_3}}{v_0N},
\end{equation}
where \(\bar{R}_g,\bar{\mu}_1,\bar{\mu}_2,\bar{\mu}_3\) are calculated directly from the PDB structures. We calculate this on a different dataset consisting of \(\sim10\)k protein structures and find \(f_V\approx 1.44\). Including this factor, Eq.\eqref{eq:rg1} becomes 
\begin{equation}
    R_g = f_V^{1/3}\Bigl(\tfrac{3v_0}{4\pi\,5^{3/2}}\Bigr)^{1/3} N^{1/3}
        \bigl(\mu_1\mu_2\mu_3\bigr)^{-1/6}.
\label{eq:rg2}
\end{equation}
which is how we scale \(R_g\) to physical sizes, and we can further use Eq.\eqref{eq:semiax} to calculate the lengths of the axes.

\section*{Data availability} 
The protein structures used in this study are publicly available from the Protein Data Bank (PDB)~\cite{rcsb}. The list of PDB identifiers, generated ion maps, structural descriptors, and the dataset splits used for testing are available from the Coherent X-ray Imaging Data Bank (CXIDB), deposition ID 251~\cite{cxidb251}. The source code used for the Coulomb-explosion simulations and data analysis is publicly available on GitHub~\cite{mc-md}. Additional data supporting the findings of this study are available from the corresponding authors upon reasonable request.

\section*{Acknowledgements}
Project grants from the Swedish Research Council (2018-00740, 2019-03935, 2023-03900) and the Helmholtz Association through the Center for Free-Electron Laser Science at DESY are acknowledged. 
CC acknowledges support from a Röntgen Ångström Cluster grant provided by the Swedish Research Council and the Bundesministerium für Bildung und Forschung (2021-05988). The computations were enabled by resources in projects NAISS 2025/5-375, 2025/22-229, 2026/4-258 provided by the National Academic Infrastructure for Supercomputing in Sweden (NAISS), funded by the Swedish Research Council through grant agreement no. 2022-06725. Computations related to the electron density analysis were performed on the Davinci computer cluster at the Laboratory of Molecular Biophysics, Uppsala University. Molecular graphics performed with UCSF ChimeraX, developed by the Resource for Biocomputing, Visualization, and Informatics at the University of California, San Francisco, with the support of the National Institutes of Health R01-GM129325 and the Office of Cyber Infrastructure and Computational Biology, National Institute of Allergy and Infectious Diseases.
\\

\section*{Author contributions statement}
A.B.: Conceptualization, Design, Simulations, Analysis, Interpretation, Writing (first draft \& editing). T.A.: Conceptualization, Design, Simulations, Analysis, Interpretation, Writing (first draft \& editing).  C.C.: Conceptualization, Design,  Interpretation, Writing.
N.T.: Conceptualization, Design,  Interpretation, Writing. 
All authors reviewed the manuscript.

\bibliography{references}

\end{document}